\documentclass[druk]{appolb}

\headauthor{J. Lang, W. Kotlarski}

\usepackage{graphicx}
\usepackage{subcaption}
\usepackage{xcolor}
\usepackage{amsmath}

\newcommand{\mytitle}{Towards automatizing Higgs decays in BSM models at one-loop in the decoupling renormalization scheme}

\usepackage[
  pdftitle={\mytitle},
  pdfauthor={Jonas Lang, Wojciech Kotlarski},
  pdfkeywords={},
  bookmarks=true, linktocpage, colorlinks=true, allbordercolors=white,
  allcolors=blue]{hyperref}

\usepackage[compat=1.1.0]{tikz-feynman}

\usepackage[    
  backend=biber,    
  bibencoding=utf8,    
  sorting=none,    
  url=true,    
  doi=false,    
  style=phys,    
  biblabel=brackets,    
  subentry=false,    
  eprint=true,    
  maxbibnames=5    
]{biblatex}
\begin{document}
\title{\mytitle\thanks{Presented by J. Lang at the 31st Cracow Epiphany Conference on the \textit{Recent LHC Results}, Krak\'ow, Poland, 13-17 January, 2025.}%
}
\author{Jonas Lang, Wojciech Kotlarski
\address{National Centre for Nuclear Research, Pasteura 7, 02-093 Warsaw, Poland}
}

\maketitle

\begin{abstract}
We propose the use of a decoupling renormalization scheme in the calculation of NLO corrections to SM-like Higgs boson decays in the beyond the Standard Model models. 
The advantage of this particular scheme is its decoupling property in the presence of a heavy BSM physics and the possibility to directly include known higher-order Standard Model corrections.
We illustrate the use of this scheme by analysing the $h \to \mu^+ \mu^-$ decay in a simple Standard Model extension by an $S_1$ leptoquark.
\end{abstract}

\section{Introduction}

Since the discovery of the Higgs boson and with recent upgrades of the LHC, the electroweak (EW) sector of the Standard Model has become a high-precision testing ground for the beyond the Standard Model (BSM) physics \cite{collaboration_observation_2012}.
With the upcoming High-Luminosity LHC upgrade, we expect measurements of electroweak observables within a few percent accuracy, even for branching ratios of the Higgs boson to light fermions \cite{atlas_measurements_2016, cepeda_higgs_2019}.
This increased precision will give a unique opportunity to probe BSM physics, including BSM Higgs sectors. 
However, the data obtained so far does not show any deviations from the SM and, in particular, confirm that the Higgs boson is SM-like.
If one wants to use the Higgs boson as a probe of New Physics, this calls for an improved accuracy of theoretical predictions.
This can be broadly achieved in two ways, by performing higher-order calculations in either the effective field theory (EFT) parametrization \cite{pomarol_higgs_2014} or directly in the BSM theory.
While the first approach requires a large separation of the SM and BSM scales, the second one works regardless of that and allows to study models with at least some light BSM states that show some promise for discovery.

A lot of work has been done in the past to calculate QCD and EW corrections to Higgs observables in the SM \cite{bahl_precision_2019}.
The go-to renormalization schemes for higher-order calculations are the $\overline{\text{MS}}$ and the on-shell schemes.
The problem is that on-shell schemes are often model-specific, while $\overline{\text{MS}}$ becomes unreliable in the case of high-mass scale, which is where the usual advantage of the aforementioned EFTs over direct calculation enters the game.

To circumvent this issue, we explore the possibility of an application of decoupling renormalization scheme in automatized BSM calculations.
Ultimately, the discussed approach will be implemented in \texttt{FlexibleDecay}, a submodule of the spectrum-generator generator \texttt{FlexibleSUSY} \cite{athron_flexiblesusy_2015,athron_flexiblesusy_2018,athron_flexibledecay_2023}.

\section{Decoupling renormalization scheme for higher-order BSM corrections}

While there are infinitely many possible renormalization schemes, usually higher-order SM calculations are performed in the $\overline{\text{MS}}$ or the on-shell scheme.
The $\overline{\text{MS}}$ scheme allows for easy analytic calculations, at the cost of introducing large corrections in the case of a heavy BSM sector and because of that can introduce large uncertainties.
The on-shell scheme, on the other hand, does not exhibit this behaviour but is usually model- and process-specific, and hard to apply in an automatised fashion.

The proposed here decoupling renormalization scheme combines advantages of both schemes. It is not plagued by large, unphysical BSM contributions and separates BSM and SM sectors in such a way that already known higher-order corrections from the pure SM can be easily incorporated.

The scheme is defined as follows.
For some parameter $P$ that exists both in the SM and BSM theory, the decoupling scheme is defined by the renormalization condition
\begin{equation}
    P^\text{dec}_\text{BSM} = P^{\overline{\text{MS}}}_\text{SM}.
    \label{eq:renormalization_condition}
\end{equation}
Here, the exponent signifies that the parameter is renormalized in the decoupling or the $\overline{\text{MS}}$ scheme, while subscript labels the theory from which it comes.
We can calculate an explicit expression for the renormalization constant $\delta P^\text{dec}_\text{BSM}$.
To do so, we parametrize the bare parameter $P_0$ as
\begin{align}
    P_0 &= P^\text{dec}_\text{BSM} + \delta P^\text{dec}_\text{BSM} = P^\text{OS}_\text{BSM} + \delta P^\text{OS}_\text{BSM} , \\
    P_0 &= P^{\overline{\text{MS}}}_\text{SM} + \delta P^{\overline{\text{MS}}}_\text{SM} = P^\text{OS}_\text{SM} + \delta P^\text{OS}_\text{SM} .
\end{align}
Using the renormalization condition and the requirement that we want to predict the same values for the on-shell renormalized parameters in both theories, we subtract both equations and obtain
\begin{equation}
    \delta P^\text{dec}_\text{BSM} = \delta P^{\overline{\text{MS}}}_\text{SM} + \delta P^\text{OS}_\text{BSM} - \delta P^\text{OS}_\text{SM} .
    \label{eq:MasterEquation}
\end{equation}
The physical significance of Eq.~\eqref{eq:MasterEquation} is captured by the division of $\delta P^\text{dec}_\text{BSM}$ into a renormalization constant $\delta P^{\overline{\text{MS}}}_\text{SM}$ that holds the SM $\overline{\text{MS}}$ contributions and the piece $\delta P^\text{OS}_\text{BSM} - \delta P^\text{OS}_\text{SM}$, the difference of on-shell renormalization constants between BSM and SM theories.
Within the first term we can add the higher-order SM corrections while the difference is free of any logarithmically growing BSM contributions.

\section{Example application: $S_1$ leptoquark}
\begin{figure}
    \centering
    \begin{subfigure}{0.45\textwidth}
    \centering
        \includegraphics[width=0.9\textwidth]{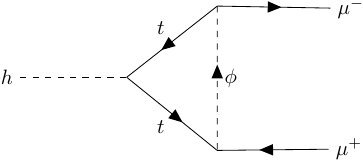}
    \end{subfigure}
    \begin{subfigure}{0.45\textwidth}
    \centering
        \includegraphics[width=0.9\textwidth]{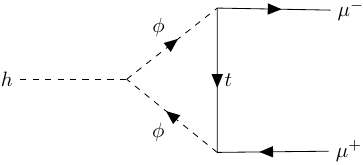}
    \end{subfigure}
    \caption{
    Leptoquark contribution to the $h \rightarrow \mu^+ \mu^-$ decay.
    By power counting, the left diagram is UV divergent, while the right one is not.}
    \label{fig:higgsDecay_FFS}
\end{figure}
Equation~\eqref{eq:MasterEquation} is the master equation for the calculation of one-loop corrections in the decoupling scheme.
To see it in action, we calculate the decay width of a Higgs boson into a lepton and anti-lepton pair in the $S_1$-leptoquark model.
While current mass limits on leptoquarks make an EFT approach valid and actually much easier, this model features several properties which make it a simple yet illustrative example.

In the $S_1$-leptoquark model, the SM gets extended by the leptoquark $\phi$ which transforms as $\left(3, 1, -\tfrac{1}{3}\right)$ under the SM gauge groups $SU(3)_c \times SU(2)_L \times U(1)_Y$. 
The SM Lagrangian gets extended in the Higgs sector by
\begin{equation}
    \mathcal{L}_H \supset -g_{H\phi} (H^\dagger H) (\phi^\dagger \phi),
\end{equation}
and in the Yukawa sector by
\begin{equation}
    \mathcal{L}_Y \supset Y^{LL}_{i j} (\overline{Q^C_i}^T \, i \sigma^2 \, L_j)\phi^\dagger + Y^{RR}_{i j} (\overline{q^C_{u,i}} \, l_{e, j}) \phi^\dagger + \text{h.c.}
\end{equation}
Here the $C$ indicates the charge conjugation operator, $Q$ and $q_u$ indicate the left-handed quark doublet and the right-handed up-type quark singlet, respectively, $L$ and $l_e$ are analogues fields in the lepton sector, and $H$ is the Higgs doublet.
For now, we are interested in the decay of a Higgs particle into a lepton and an anti-lepton.
At the one-loop level, there are two Feynman diagrams that contain leptoquarks, which are depicted in Figure~\ref{fig:higgsDecay_FFS}.
When analysing the model in the decoupling scheme, we need to look at two behaviours of these diagrams. 
First, by power-counting, we get that the left diagram is UV divergent, while the second diagram is not. 
The second point concerns the BSM scale, in this case, the leptoquark mass. 
Increasing the mass also increases the weight of these diagrams. 
We see this when we look at the amplitude of this decay
\begin{equation}
    \begin{split}
        i \mathcal{A} &= F_L P_L + F_R P_R \\
        &= \frac{3 m_{t}}{16 \pi^2 v} \left\{\left[B_0(m^2_{\mu}, m^2_\phi, m^2_{\mu}) + \dots \right] (Y^{RR\dagger})_{2 3} Y^{LL}_{3 2} + \dots \right\} P_L \\
        & \quad+ \frac{3 m_{t}}{16 \pi^2 v} \left\{\left[B_0\left(m^2_{\mu}, m^2_\phi, m^2_{\mu}\right) + \dots \right] (Y^{LL\dagger})_{2 3} Y^{RR}_{3 2} + \dots \right\} P_R.
    \end{split}
\end{equation}
Here, $m_{t}$ is the top quark mass, $m_{\mu}$ is the muon mass, $v$ is the vacuum expectation value of the Higgs field, $B_0$ is the Passarino--Veltman (PV) function and dots stand for contributions proportional to other PV functions. Looking at the analytic expression for the $B_0$ function
\begin{equation}
    B_0(p, m_1, m_2) = \frac{1}{\epsilon} - \log\left(\frac{m^2_1}{\mu^2}\right) + 1 - \int^1_0 dx \log\left(1 + \frac{1-x}{x} \frac{p^2}{m^2_1} - (1 - x) \frac{m^2_2}{m^2_1} \right),
\end{equation}
\begin{figure}[tb]
    \centerline{%
        \includegraphics[width=.85\textwidth]{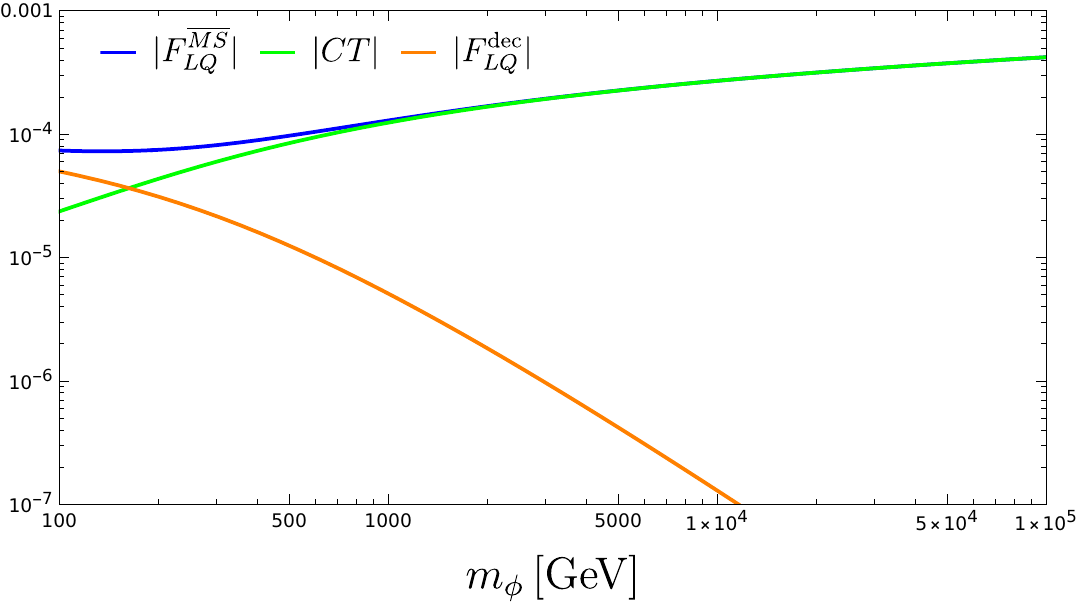}
    }
    \caption{Leptoquark contribution to the $\overline{\text{MS}}$ renormalized form factors (blue), the decoupling-scheme counter term (green) and the form factor renormalized in the decoupling scheme (orange).
    The $\overline{\text{MS}}$ result without a decoupling counter term exhibits a logarithmic growth with the increasing BSM scale.}
    \label{Fig:behavior_formFactors}
\end{figure}
we see the logarithmic increase with the leptoquark mass, which eventually leads to a divergence for an infinitely heavy leptoquark.
\begin{figure}[tb]
    \centerline{
        \includegraphics[width=.85\textwidth]{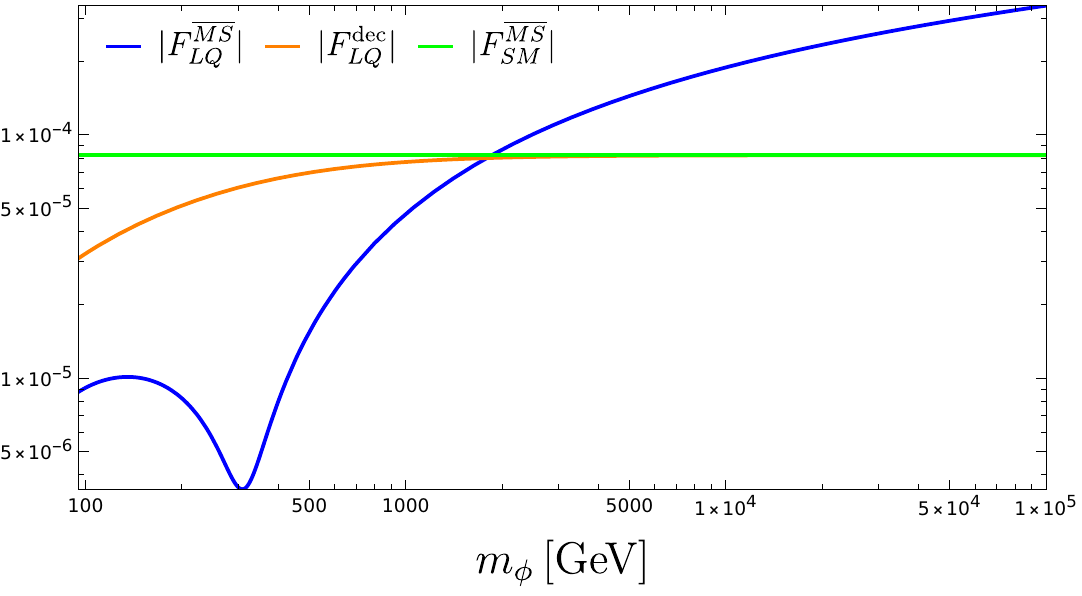}
    }
    \caption{Complete form factors (including all SM and BSM contributions) for the $h \to \mu^+ \mu^-$ decay. The blue line shows the $\overline{\text{MS}}$ renormalized form factors, the orange line the corresponding form factor in the decoupling scheme and the green line the pure SM $\overline{\text{MS}}$ result. The blue line features the expected logarithmic growth with the mass scale of the BSM sector. This is addressed in the decoupling scheme, where the renormalized form factor approaches the SM result.}
    \label{Fig:behavior_formFactors_BSMpSM}
\end{figure}
This behaviour is the root of possible large uncertainties in other renormalization schemes. 
If we now include the counterterm for the Higgs decay in the decoupling scheme, given by Eq.~\eqref{eq:MasterEquation}, we find that the $B_0$ contribution is completely subtracted and the renormalized amplitude decreases with increasing leptoquark mass. 
This behaviour is depicted in Figure~\ref{Fig:behavior_formFactors}. 
The plots are created by fixing all the input parameters, that is the masses and Higgs couplings, and varying only the leptoquark mass. 
This captures the general behaviour of the decoupling renormalization scheme.
The blue line shows the $\overline{\text{MS}}$ renormalized form factor which clearly exhibits the logarithmic growth for large leptoquark masses.
The counterterm, the green line, has the same behaviour.
Adding both together results in the orange line, which represents the form factor renormalized in the decoupling scheme, that tends to 0 in the $m_\phi \to \infty$ limit.

As the next step, we can include the remaining contributions to the Higgs particle decaying to muons. The form factors in this case are depicted in Figure~\ref{Fig:behavior_formFactors_BSMpSM}.
The green line shows the $\overline{\text{MS}}$ renormalized SM contributions.
The blue line shows the $\overline{\text{MS}}$ renormalized BSM form factors and the orange line form factors renormalized in the decoupling scheme, where the renormalized form factor approaches the SM result.

\section{Summary}
We propose the use of a decoupling renormalization scheme for the calculation of one-loop BSM corrections to SM-like Higgs boson decays.
This scheme's renormalization constants consist of two pieces.
First, there is a part that captures the $\frac{1}{\epsilon}$ divergencies of the SM contribution.
The second part is the difference between the on-shell renormalization constants in the full theory and in the SM.
This last part is by construction UV-finite and free of large logs in the presence of heavy BSM physics.
This makes this scheme reliable in parameter spaces motivated in light of the non-observation of any light BSM physics, as well as particularly well-suited for automatization of the calculation of Higgs boson decays in arbitrary BSM models.

We show how this renormalization scheme works in practice by calculating the Higgs decay to muons in the $S_1$-leptoquark extension of the SM.
We prove by direct calculation that the renormalized $h \to \mu^+\mu^-$ form factor has a decoupling property and converges to the SM result in the limit of a large leptoquark mass.
\\

Work supported by the National Science Centre (NCN), Poland, grant 2022/\allowbreak47/\allowbreak D/\allowbreak ST2/\allowbreak03087.

\printbibliography

\end{document}